\documentstyle[prl,aps,array,epsf,twocolumn,psfig]{revtex}

\def\be{\begin{equation}}
\def\ee{\end{equation}}
\def\ba{\begin{array}}
\def\ea{\end{array}}
\def\hb{\hfill\break}
\def\tetris{T{\sc etris}~}

\begin{document}

\title{Criticality of the ``critical state'' of granular media:\\
Dilatancy angle in the \tetris model}

\author{Marina Piccioni$^{(1,2)}$, Vittorio Loreto$^{(1)}$ and St\'ephane
Roux$^{(3)}$}

\address{(1): Laboratoire de Physique et M\'ecanique des Milieux
H\'et\'erog\`enes,\\
Ecole Sup\'erieure de Physique et Chimie Industrielles,\\ 
10 rue Vauquelin, 75231 Paris Cedex 05, France.\\
(2): Dipartimento di Scienze Fisiche and Istituto Nazionale di 
Fisica della Materia,\\ 
Universit\'a di Napoli, Mostra d'Oltremare, Pad. 19, 80125 Napoli,
Italy.\\
(3): Laboratoire Surface du Verre et Interfaces,\\
Unit\'e Mixte de Recherche CNRS/Saint-Gobain,\\
39 Quai Lucien Lefranc, 93303 Aubervilliers Cedex, France.\\}
\date{\today}
\wideabs{

\maketitle
\begin{abstract}
The dilatancy angle describes the propensity of a granular medium to
dilate under an applied shear.  Using a simple spin model (the ``\tetris'' 
model)  which accounts for geometrical ``frustration'' effects, we
study such a dilatancy angle as a function of density.  An exact mapping
can be drawn with a directed percolation process which proves that there
exists a critical density $\rho_c$ above which the system expands and
below which it contracts under shear.  When applied to packings
constructed by a random deposition under gravity, the dilatancy angle is
shown to be strongly anisotropic, and it constitutes an efficient tool
to characterize the texture of the medium. 
\end{abstract}}

\section{Introduction}

Granular materials \cite{grain} give rise to a number 
of original phenomena, which
mostly result from their peculiar rheological behavior.  Even using the
most simple description of the grains (rigid equal-sized spherical
particles) a granular system displays a rather complex behavior which
shows that the origin of this rheology has to be found at the level of
the geometrical arrangement of the grains. 

Guided by these considerations, models have been proposed to account for
the geometrical constraints of assemblies of hard-core particles
\cite{Tetris,RTM,degennes,Ben-Naim,NCH,baxter,Lintz}.  
The motivation of these models is not to reproduce faithfully the local
details of granular media, but rather to show that simple geometrical
constraints can reproduce under coarse-graining some features observed
in real granular media.  Along these lines, one of the most impressive
examples is the ``Tetris'' model \cite{Tetris,RTM} which, in its simplest version, 
is basically a spin model with
only hard core repulsion interactions.  This model has been introduced
in order to discuss the slow kinetics of the compaction of granular
media under vibrations.  In spite of the simplicity of the definition of
the model, the kinetics of compaction has been shown to display a very
close resemblance to most of the experimentally observed features of
compaction\cite{Knight} and segregation\cite{segtet}.

\begin{figure}
\centerline{\epsfxsize=.9\hsize \epsffile{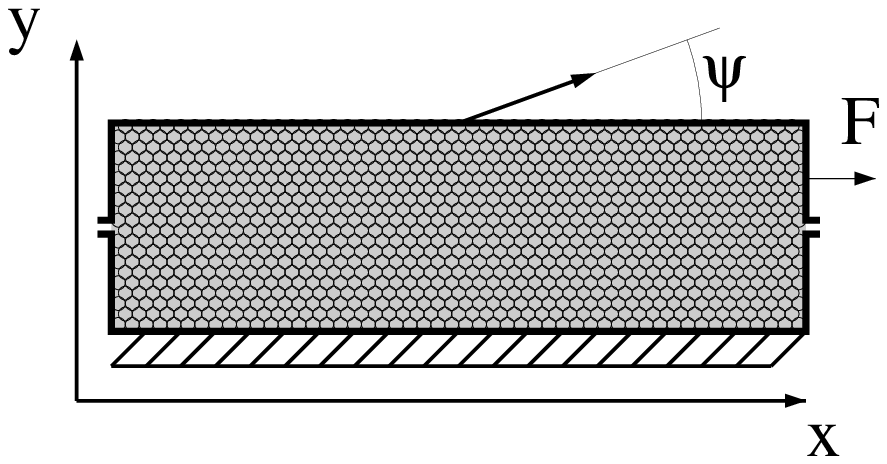}}
\caption{}
\label{uno}
\end{figure}

Our aim is here to consider again the \tetris model and to focus on a
basic property of the quasi-static shearing of a granular assembly.  It
is well known since Reynolds that dense granular media have to dilate in
order to accommodate a shear\cite{rey}, whereas loose systems contract.  This
observation is important since it gives access to one of the basic
ingredients (the direction of the plastic strain rate) necessary to
describe the mechanical behavior in continuum modeling.  
The dilatancy angle is defined as the ratio of the rate of volume
increase to the rate of shearing.  Denoting with $\varepsilon_{xy}$
the component $xy$ of the strain tensor $\varepsilon$, Fig.(\ref{uno}) illustrates an 
experiment where a shear rate $\dot\varepsilon_{xy}$ is imposed together
with a zero longitudinal strain rate $\dot\varepsilon_{xx}=0$, and the
volumetric strain rate (here vertical expansion) $\dot\varepsilon_{yy}$ 
is measured.  The
direction of the velocity of the upper wall makes an angle $\psi$ with
respect to the horizontal direction.  This angle is called the dilatancy
angle, $\psi$.  In this particular geometry we have 
\be
\tan(\psi)={\dot \varepsilon_{yy}\over\dot\varepsilon_{xy}} 
\ee
More generally, the tangent of the dilation angle is the ratio of
the volumetric strain rate (${\rm tr}(\dot \varepsilon)$) to the
deviatoric part of the strain rate.

Numerous experimental studies have confirmed the validity of such a
behavior, and have lead to extensions such as what is known in soil
mechanics as the ``critical state'' concept\cite{soilmech}.   
Assuming that the incremental (tangent) mechanical behavior can be parametrized 
using only the density of the
medium, $\rho$, a loose medium will tend under continuous shear towards
a state such that no more contraction takes place, i.e. it will assume 
asymptotically a density $\rho_c$ such that $\psi(\rho_c)=0$.  This
state is by definition the ``critical state''.  Conversely, if the
strain were homogeneous, a dense granular media would dilate until it
reached the critical state density $\rho_c$.  However, for dense media,
the strain may be localized in a narrow shear band which may allow a
further shearing without any more volume change so that the mean density
may remain at a value somewhat higher than the critical value.  Recent
triaxial tests \cite{triaxial} in a scanner apparatus have however shown that in the
shear bands the density of the medium was quite comparable to the
critical density, thus providing further evidence for the validity of
the critical state concept.   

The word ``critical'' used in this context has become the classical
terminology, but it has no a priori relation with any kind of {\em
critical} phenomenon in the statistical physics vocabulary
\cite{crit-phen}.  
One of the results presented in this article is to show that indeed the critical
state of soil mechanics is also a critical point in the sense of 
phase transitions, for the \tetris model considered here.

\section{Model and definition of dilatancy}

A group of lattice gas models in which the main ingredient is 
the geometrical
frustration has been introduced recently under the name \tetris
\cite{Tetris,RTM}.

\begin{figure}
\centerline{\epsfxsize=.7\hsize \epsffile{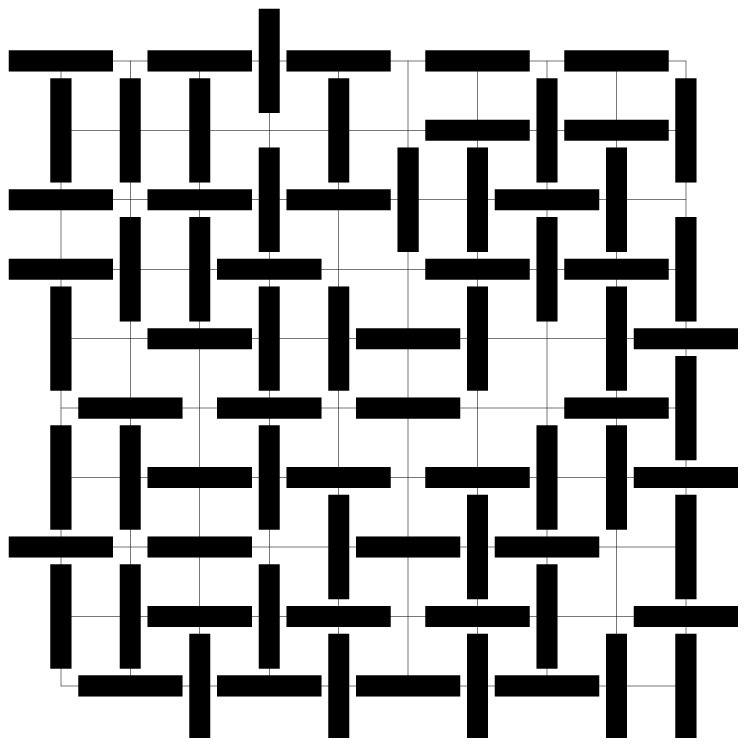}}
\caption{}
\label{due}
\end{figure}

The \tetris model is a simple lattice model in which the sites of 
a square lattice can be occupied by (in its simplest version) 
a single type of rectangular shaped particle with only two possible 
orientations along the principal axis of the underlying lattice.
A hard core repulsion between particles is considered so that two 
particles cannot overlap.  This forbids in particular that two nearest 
neighbor sites could be both occupied by particles aligned with the 
inter-site vector.  An illustration of a typical admissible 
configuration is shown schematically in Fig.(\ref{due}).
More generally one can consider particles that move on a
lattice and present randomly chosen shapes and sizes\cite{RTM}. 
The interactions in the system obey to the general rule that one cannot
have particle overlaps. The interactions are not spatially quenched 
but are determined in a self-consistent way by the local
arrangements of the particles.

The definition of the dilation angle as sketched in Fig.(\ref{uno}) 
is difficult to implement in practice in the \tetris model due to the underlying
lattice structure which defines the geometric constraints only for
particles on the lattice sites, and not in the continuum.

\begin{figure}
\centerline{\epsfxsize=.7\hsize \epsffile{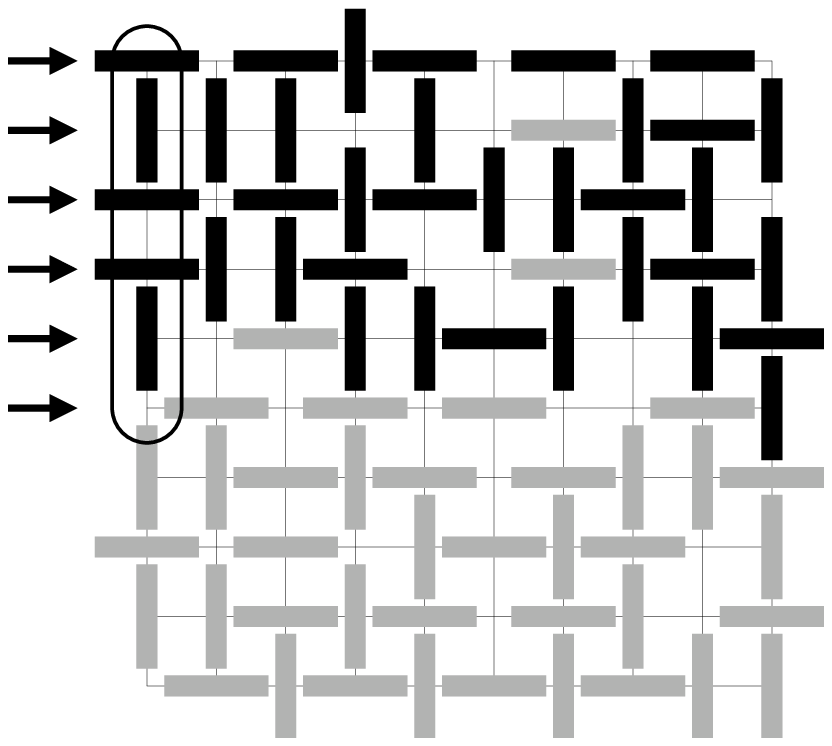}}
\caption{}
\label{tre}
\end{figure}

We may however circumvent this difficulty through the following
construction illustrated in Fig.(\ref{tre}).  We consider a semi-infinite line
starting at the origin and oriented along one of the four cardinal
directions.  This line is (and all the sites attached to it are) pushed
in one of the principal directions of the underlying square lattice by
one lattice constant.  In the following, we will consider only a
displacement perpendicular to the line, although a parallel displacement
may also be considered.  As this set of particles is moved, all other
particles which may overlap with them are also translated with the same
displacement. In this way, we determine the set $\cal D$ of particles
which moves.  In the sequel, we will show that this domain is nothing
but a directed percolation cluster \cite{DP} grown from the line.  
Anticipating on the following, the mean shape of $\cal D$ will be shown 
to be a wedge limited by a generally rough boundary whose mean orientation 
forms an angle $\psi$ with the direction of motion.  The angle $\psi$ can be
shown to be exactly equal to the dilatancy angle as defined previously.

Exploiting the non-overlap constraint, we may simply determine the
rule for constructing the domain $\cal D$.   Let us choose the
particular case of a displacement in the direction $(1,0)$, and consider
a non-empty site $(i,j)$ which is displaced.  The particles which may
have to be displaced together with site $(i,j)$ can be identified
easily:\hb
-If the particle in $(i,j)$ is horizontal:
\hb $\bullet$
$(i+1,j)$ if the site is occupied by a particle with any orientation.
\hb $\bullet$
$(i+2,j)$ if the site is occupied by a horizontal particle.
\hb
- If the particle in $(i,j)$ is vertical:
\hb $\bullet$
$(i+1,j)$ if the site is occupied by a particle with any orientation.
\hb $\bullet$
$(i+1,j\pm 1)$ if the site is occupied by a vertical particle.

Using these rules, it is straightforward to identify the cluster of
particles $\cal D$.  The model thus appears to be a directed percolation
problem with a mixed site/bond local formulation.  Thus unless long
range correlations are induced by the construction of the packing, the
resulting problem will belong to the universality class of directed
percolation.  The density of particles, $p \in [0,1]$, in the lattice plays 
the role of the site or bond presence probability, i.e. the control parameter of the
transition.

Let us recall, for sake of clarity, some properties of the two-dimensional
directed percolation. For $p < p_c$ (where $p_c$ is the directed percolation
threshold), a typical connected cluster extends over a distance of the order of
$\xi_{/\!\!/}$ in the parallel direction (the preferential direction) and a distance
$\xi_{\perp}$ in the perpendicular direction. For $p > p_c$ there appears
a directed percolating cluster which extends over the whole system. This cluster
possesses a network of nodes and compartments. Each compartment has an 
anisotropic shape similar to the connected cluster below $p_c$, characterized by 
$\xi_{/\!\!/}$ in the parallel direction and $\xi_{\perp}$ in the perpendicular direction.
On both sides of the percolation transition, the two lengths present the power-law behavior
$\xi_{/\!\!/} \sim | p-p_c|^{-\nu_{/\!\!/}}$ and 
$\xi_{\perp} \sim | p-p_c|^{-\nu_{\perp}}$.

\section{Monocrystal}

\begin{figure}
\centerline{\epsfxsize=.7\hsize \epsffile{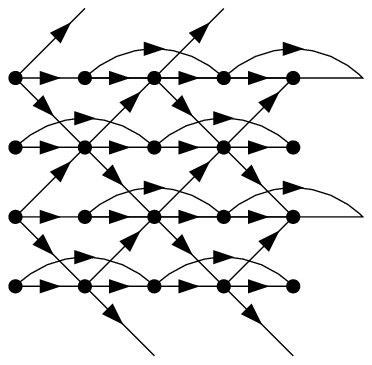}}
\caption{}
\label{quattro}
\end{figure}

Let us first examine a simple geometrical packing. There exist (two) special
ordered configurations of particles such that the density can reach
unity (one particle per site).  This corresponds to a perfect
staggered distribution of particle orientations.  Thus a simple
way of continuously tuning the density is to randomly dilute one of
these perfectly ordered states.  In this case, if a site is occupied by
a particle, its orientation is prescribed.  Therefore the above rules
can be easily reformulated as a simple directed site percolation problem in a
lattice having a particular distribution of bonds (up to second
neighbors).
Fig.(\ref{quattro}) illustrates the specific distribution of bonds corresponding to
such an ordered state.

For $p=1$, suppose that the initial seed is $(0,j)$ for $j\ge 0$ and 
this line is pushed in the $x$ direction.  Then the infinite cluster is
the set of sites $(i,j)$ such that $j\ge -i$, for a vertical spin at the
origin.  Thus moving the semi-infinite line (seed) introduces vacancies
in the lattice which was initially fully occupied.  The system dilates
and its dilation angle is $\psi_1=\pi/4$.

As $p$ is reduced, the orientation of the boundary changes up to the
stage where it becomes parallel to the $x$ axis for $p=p_R$.  At this point the
dilatancy is zero. A motion is possible without changing the volume.
This point corresponds precisely to the directed percolation threshold 
(using the precise rules defined above).

From the theory of directed percolation, we can directly
conclude that the behavior of the dilatancy angle $\psi$ in the
vicinity of $p_R$ obeys
\be\label{eq:critbeha}
\tan(\psi)\propto (p-p_R)^{\nu_{/\!\!/}-\nu_{\perp}}
\label{tan}
\ee
where the correlation length exponents are $\nu_{/\!\!/}\approx 1.732$ and 
$\nu_{\perp}\approx 1.096$ independently of the lattice used.

\begin{figure}
\centerline{\epsfxsize=.7\hsize \epsffile{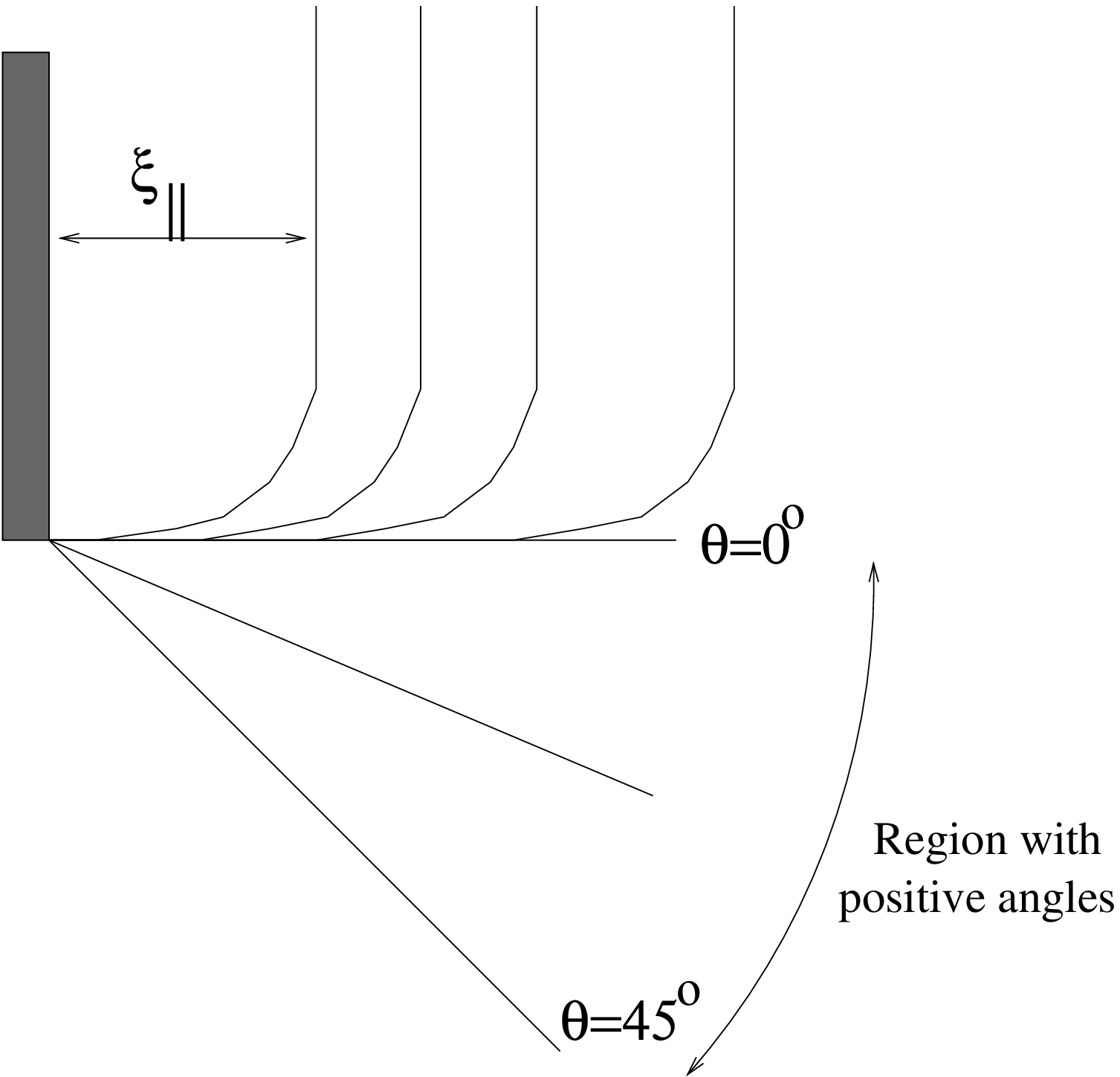}}
\caption{}
\label{cinque}
\end{figure}

A further decrease of $p$ leads to a subcritical regime where only a
finite cluster is connected to the initial seed.  Only a finite
layer of thickness $\xi_{/\!\!/}\propto (p_R-p)^{-\nu_{/\!\!/}}$ along
the $y$-axis is mobilized. This means that it not possible 
to define in the same way the dilation angle for $p <p_R$
(negative angles).  What happens in practice is that for
$p <p_R$ the shearing produces a compaction of the system in front 
of the semi-infinite line pushing the system.
Fig.(\ref{cinque}) summarizes schematically the situation for all the values
of $p$. The horizontal line corresponds to $p= p_R$ and a zero 
dilation angle.

We performed numerical simulations of this problem using a transfer
matrix algorithm which allowed to generate system of size up to
$10^4\times3 \cdot 10^4$.  These large system sizes allowed for a very accurate
determination of the dilatancy angle as a function of the occupation
probability (density) $p$.  Fig.(\ref{cinque}) shows the boundaries of the domains
$\cal D$ for $p=0.58$, close to the directed percolation threshold $p_R$,
 and $p=0.7$.

\begin{figure}
\centerline{
      \psfig{figure=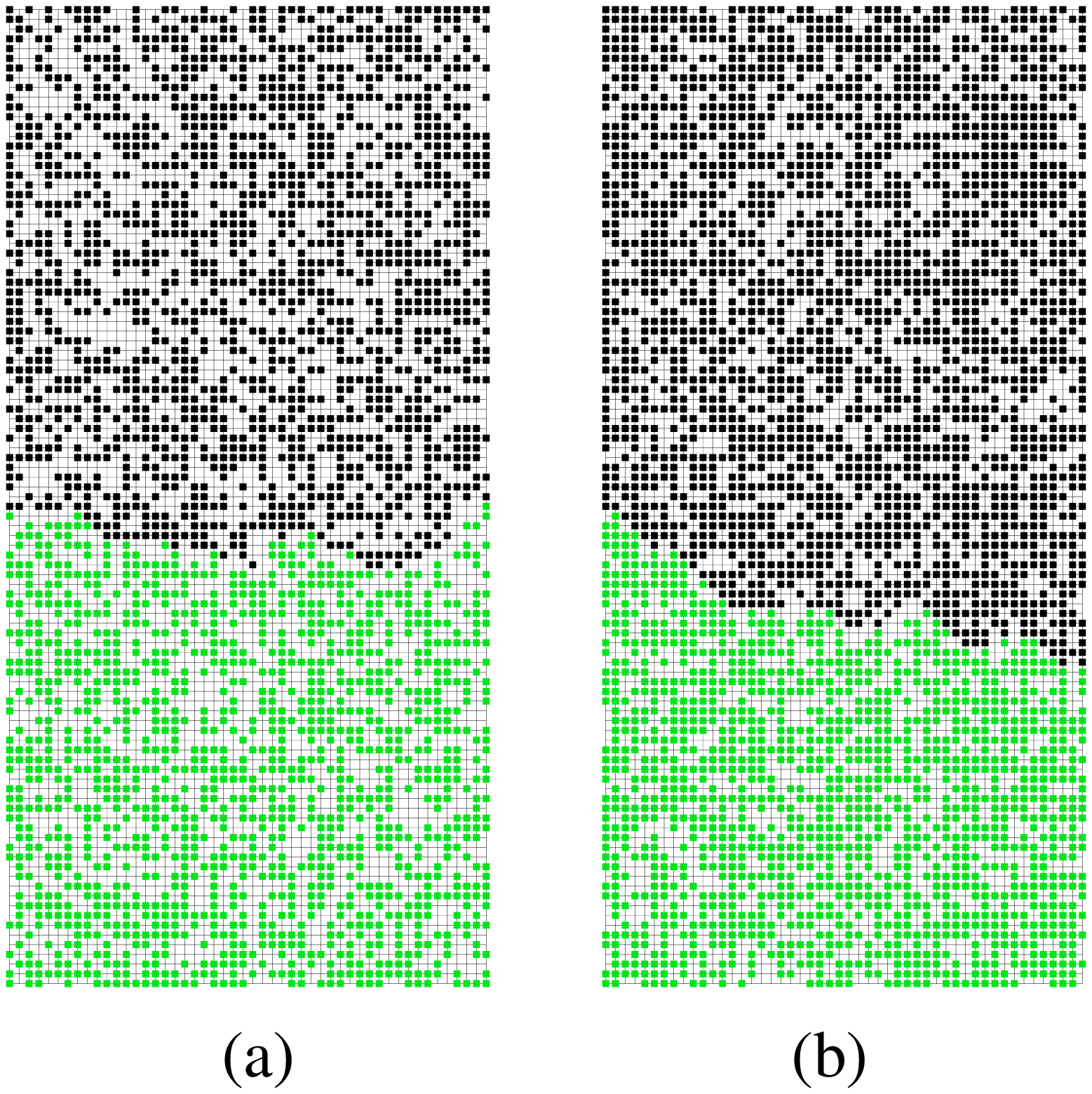,width=5cm,angle=-90}}
\caption{}
\label{sei}
\end{figure} 

\begin{figure}
\centerline{
       \psfig{figure=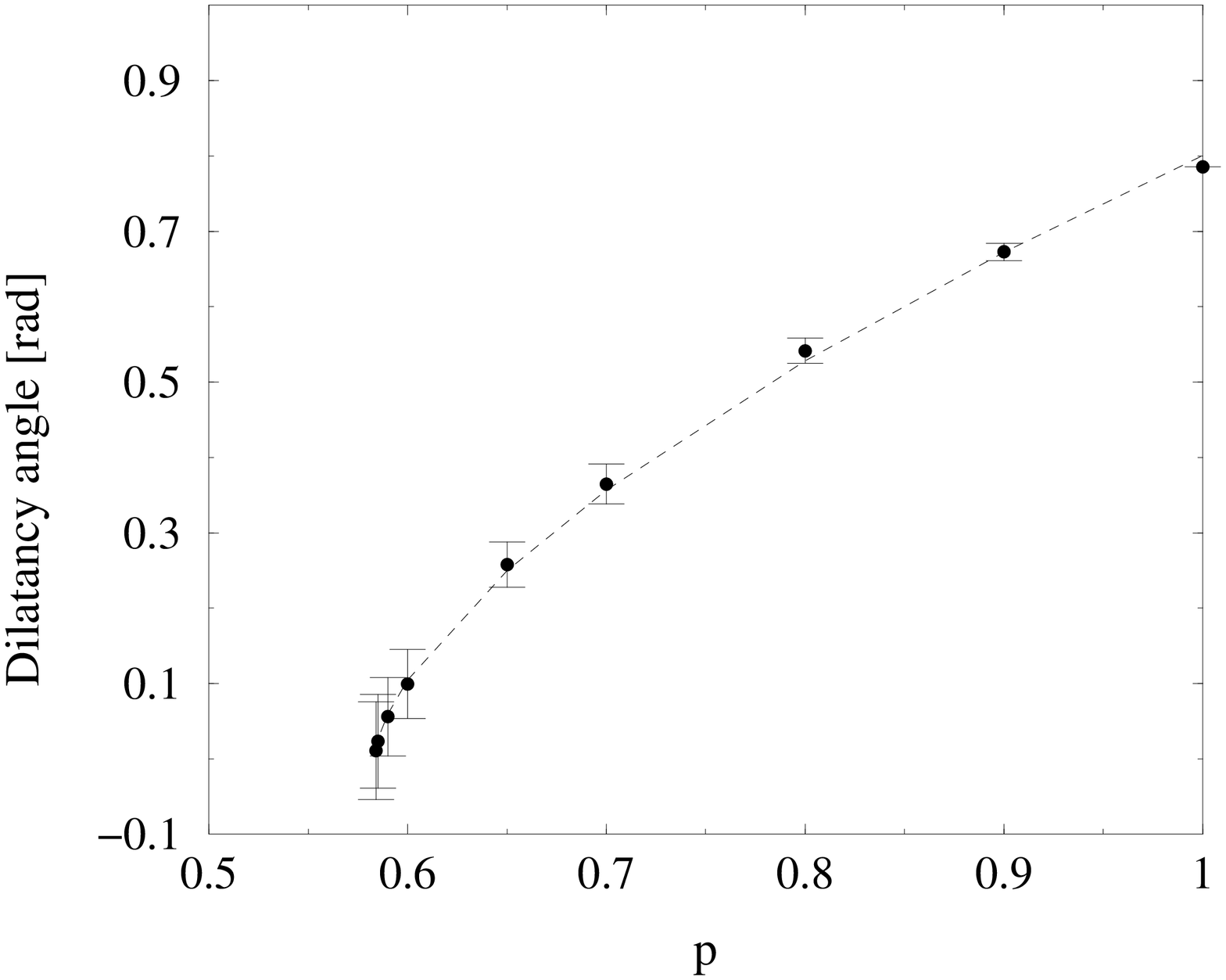,width=8cm,angle=-90}}
\caption{}
\label{sette}
\end{figure}

Fig.(\ref{sette}) shows the estimated dilatancy angle as a function of the
density of particles. Angles are evaluated on lattice of size 
$10^4 \times 3\cdot 10^4$ and are averaged over $100$ different realizations.
The  onset of dilatancy is thus estimated to be
\be
p_R=0.583\pm0.001
\ee

The singular variation of $\psi$ close to the onset of dilatancy
Eq.~\ref{eq:critbeha} has been checked to be consistent with our 
numerically determined values as shown by the dotted curve in Fig.(\ref{sette})
which corresponds to the expected critical behavior.

\section{Random sequential deposition}

It is worth emphasizing that the directed percolation problem associated
with the dilatancy angle determination is simply a site percolation
problem in the above special case where each site is assigned only one
possible orientation for the particle.  In the more general
case, the way the cluster is grown locally depends on the specific orientation
of the particle.  Thus it is a mixed site/bond percolation problem.  
Therefore, depending on the way the system has been built, the onset for
dilatancy, $p_R$, will vary.  

This is illustrated by constructing the system through a random
deposition process, i.e. differently from the above procedure.    The
algorithm used to construct the system is the following.  At each time
step, an empty site and a particle orientation are chosen at random.  If
the particle can fit on this site (without overlap with other
particles), then the site is occupied, otherwise a new random trial is
made. This is similar to the ``random sequential'' problem
often studied in the literature\cite{evans}, here adapted to the \tetris model.

This procedure  leads to a maximum density of particles around
$p_{max}\approx 0.75$ above which it becomes impossible to add
new particles. 

\begin{figure}
\centerline{
       \psfig{figure=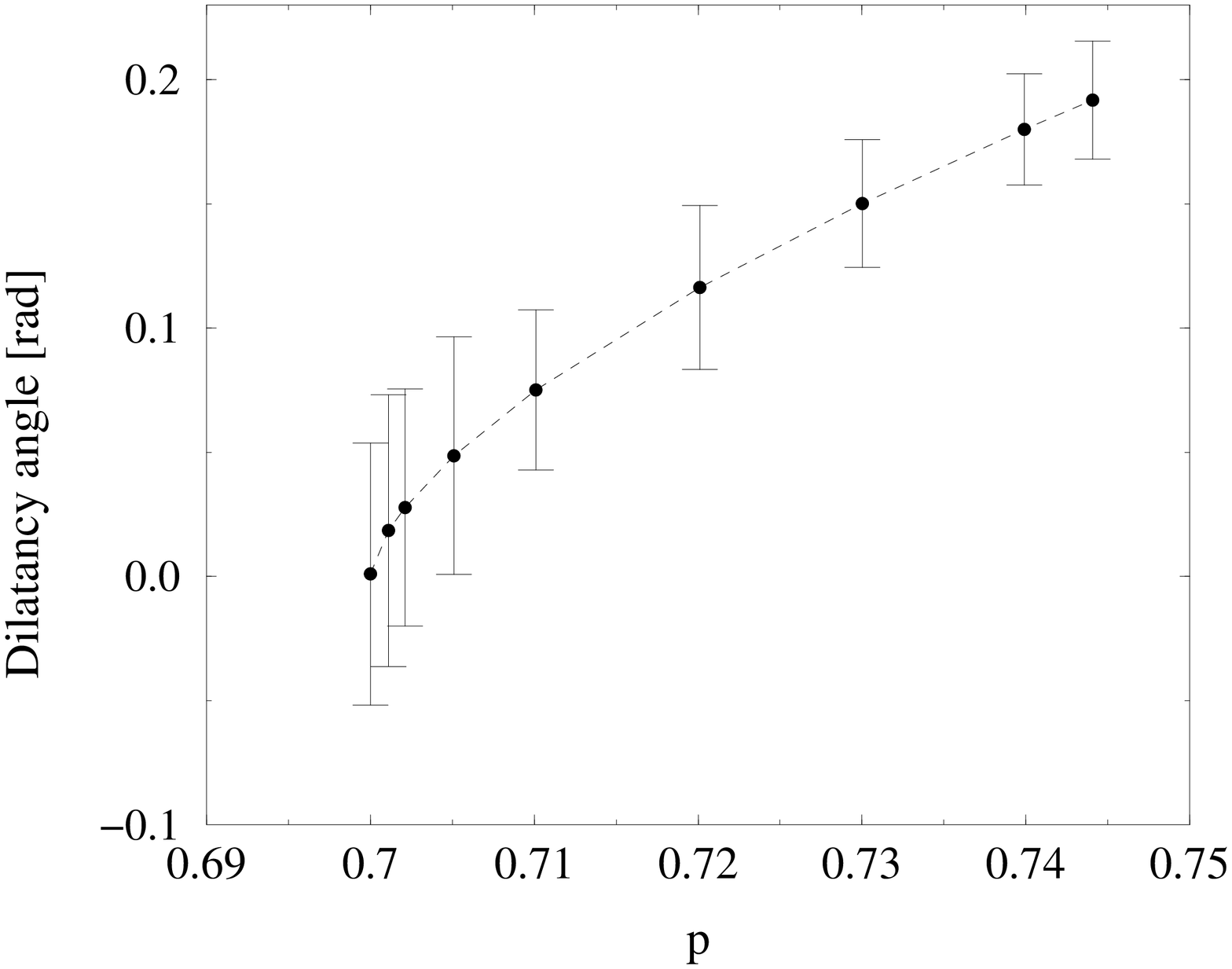,width=8cm,angle=-90}}
\caption{}
\label{otto}
\end{figure}

Differently from the previous case, in the random
sequential deposition simulations
could not have been performed using the transfer matrix algorithm and thus
we generated systems of size up to $500\times 1500$.
We studied the dilatancy angle in such systems stopping the construction
at different $p$ values, averaging for each $p$ over 
$100$ realizations.
Fig.(\ref{otto}) shows the estimated dilatancy angle which is definitely
different from the data of Fig.(\ref{sette}).  In particular the onset of
dilatancy is determined to be 
\be
p_R=0.70\pm 0.01
\ee

However, this procedure is not expected to induce long range
correlations in the particle density or orientation, and thus, we expect
that the universality class of the model remains unchanged.  In
particular, the critical behavior Eq.~\ref{eq:critbeha} is expected to
hold with the same exponents.  Although the system sizes are much
smaller in the present case, our data are consistent with such a law.

\section{Ballistic deposition under gravity}

Finally we would like to point out another property related to the
texture of the medium.  Up to now the two procedures followed to generate
the packing of particles did not single out any privileged direction.

We now construct the packing by random deposition under gravity. 
Particles with a random orientation are placed at a random $x$ position,
and large $y$.  Then the particle falls (along $-y$) down to the first site
where it hits an overlap constraint.  In this way, the packing assumes a
well defined bulk density $p\approx 0.8$.  

We used this construction procedure to generate lattices of size 
$500 \times 1500$ (averaged over 500 samples) cutting out the top part of the 
lattice which is characterized by a very wide interface and a non-constant density 
profile. On this configuration (and thus at a fixed density) 
we measured the dilatancy angle for different orientations of the imposed
displacement on the wall with respect to ``gravity''.  

Table 1 shows the resulting dilatancy angles obtained for the same density
$p=0.81$ using different constructions:
\begin{itemize}
\item the dilution of the ordered state, 

\item the sequential deposition, (in both of these cases the dilatancy
angle does not depend on the orientation of the motion). It is worth noticing
how a direct comparison between this case and the others is not possible because
with the Random Sequential Deposition one cannot obtain densities
larger that $\simeq 0.75$.

\item the ballistic deposition using a displacement along $-y$ (against
gravity), $y$ (along gravity), and $x$ (perpendicular to gravity).  In 
the latter case, we could study the problem for two orientations of the
semi-infinite line ($x=0$ and $y>0$ or $y<0$).  We checked that the
dilatancy angle was not dependent on this orientation.

\end{itemize}

The data reported in Table 1 indeed shows that the dilatancy angle can be 
dependent on the direction of the imposed displacement.   This
measurement is thus sensitive to texture effects.  As a side result, we
note that the usual characterization of the dilatancy in terms of a
single scalar (angle), albeit useful, is generally an  oversimplification
for textured media.  Indeed, a number of studies have revealed \cite{exp-dil} 
that granular media (even consisting of perfect spheres) easily develop 
a non isotropic texture as can be shown by studying the distribution of
contact normal orientations. This remark is almost obvious from a
theoretical point of view, however, few attempts have been made to
incorporate these texture effects in the dilatancy angle or even more
generally in the rheology of granular media.

\begin{table}
\begin{center}
\begin{tabular}{|c|c|c|}
\hline
Method & Orientation &$\psi$\\
\hline
\hline
Dilution&$\pm x \pm y$  & 31.0 $\pm$ 0.1\\
BDG&$-y$   & 6.6 $\pm$ 0.5\\
BDG&$\pm x$& 23.8 $\pm$ 0.5\\ 
BDG&$+y$   & 5.8 $\pm$ 0.5\\ 
\hline
\end{tabular}
\end{center}
\caption{Results for the 
dilatancy angles obtained using differently prepared samples and 
different displacement orientations at $p\approx 0.8$. 
Dilution indicates samples obtained
by diluting a perfect monocrystal (see text) to the desired density;
BDG indicates samples obtained with a Ballistic Deposition procedure 
under Gravity. We cannot compare directly in this table the results obtained
with the Random sequential deposition procedure (RSD) because,
as mentioned in the text, this procedure  leads to a maximum density 
of particles around $p_{max}\approx 0.75$ above which it becomes 
impossible to add new particles.}
\end{table}

\section{Conclusion}
We have shown that dilatancy can be precisely defined in the \tetris
model, and that it is a function of the density as it is well known for
granular media.  The onset of dilatancy, i.e. the ``critical state'' of
soil mechanics, has been shown to corresponds to a directed percolation
threshold density, hence justifying the term ``critical'' in this
expression.  Form this point of view it is important to stress 
how any comparison of our approach with real granular materials 
should be done in the neighborhood of the critical point where 
we expect a largely universal (in the sense of critical
phenomena in the statistical physics vocabulary) behavior. 
Using different lattices we expect, for instance, to recover
the same critical behavior (same exponents) but not the same values
for the critical density.
To our knowledge this is the first time that such a mapping
is proposed.  We have also shown that the dilatancy angle was not only
determined by the density but also by the packing history.  
Finally, we have shown from a simple anisotropic construction that texture affects
the dilatancy angle, even for a fixed density.

\acknowledgements

We wish to thank S. Krishnamurthy, H.J. Herrmann and F. Radjai for useful
discussions related to the issues raised in this study. 
This work has been partially supported  from the European 
Network-Fractals under contract  No. FMRXCT980183.

\newpage

\section{figure captions}
\begin{itemize}

\item[Fig.(1)] Schematic view of shearing of granular media in a shear cell. 
The upper part of the cell moves only if the medium dilates so that the
direction of the motion forms an angle $\psi$, the dilatancy angle, 
with the horizontal direction.

\item[Fig.(2)] Illustration of the \tetris model.  The sites of a square
lattice can host elongated particles shown as rectangles.  The width
and length of the particles induce geometrical frustrations.

\item[Fig.(3)] Procedure used to define the dilation angle.  All particles 
located on a semi-infinite line (the particles enclosed in the round-edge 
rectangle on the left-hand side of  the lattice) are moved by one  lattice
unit in the horizontal direction (shown by the arrows). Using the
hard-core repulsion between particles, we determine the  particles which
are pushed (shown in black) and those which may  stay in place (grey). 
For each column we consider the lowest (in general the most external)  
black site (The gray particles within the cluster of black particles
do not play any role in the determination of the dilation angle).
The curve connecting all these points defines the profile of the 
pushed region. The line connecting the first and the last points of this 
profile determines the angle, $\psi$,  with respect to the 
direction of motion. This angle provides the value of the dilation angle for 
the particular realization considered. The dilation angle is actually measured 
performing an average over a large number of realizations.

\item[Fig.(4)] Lattice over which directed site percolation is taking place. 
The arc bonds connect second neighbors along the $x$ axis (horizontal).

\item[Fig.(5)] Schematic representation of the mobilized region
in the shearing procedure. Starting from $p=1$, where one has a dilation
with an angle of $\pi/4$, the dilation angle reduces until $0$ (for $p=p_R$).
A further reduction of $p$ brings the system in a subcritical regime
where only a finite layer of thickness $\xi_{/\!\!/}\propto (p_R-p)^{-\nu_{/\!\!/}}$ 
along the $y$-axis is mobilized and the system compactifies.

\item[Fig.(6)] Shapes of the boundaries of two clusters for (a) $p=0.58$ 
(close to the threshold for a vanishing dilatancy) and (b) $p=0.7$.
The clusters mobilized are above and in both cases the line 
interpolating linearly between the first and  the last point the boundaries 
defines the dilation angle.

\item[Fig.(7)] Dilatancy angle as a function of the density $p$ in the case of
a random dilution of the perfectly ordered \tetris model. The dashed line
represents a fit obtained using Eq.~(\ref{tan}) with $p_R=0.583\pm0.001$.
The relative errors diverge at the transition.

\item[Fig.(8)] Dilatancy angle as a function of the density $p$ in the case of
a random sequential deposition. The dashed line
represents a fit obtained using Eq.~(\ref{tan}) with $p_R=0.70\pm 0.01$.
The relative errors diverge at the transition.

\end{itemize}

\end{document}